1   **Understanding thermoregulatory transitions during haemorrhage by piecewise**

2   **regression**



4   P. S. Reynolds[1] and G. S. Chiu[2, 3]






7   [1]Department of Emergency Medicine and Virginia Commonwealth University

8   Reanimation Science Center (VCURES), Virginia Commonwealth University Medical

9   Center, Richmond, VA USA;

10  [2]CSIRO Mathematics, Informatics and Statistics, Canberra, ACT Australia

11  [3]Department of Statistics and Actuarial Science, University of Waterloo, Waterloo, ON

12  Canada




14  **RUNNING HEAD**: Estimating thermoregulatory transitions

15  **KEY WORDS**: Bent cable regression, body temperature, change-point estimation




17  **Please address all correspondence to:**
18  Dr. P. S. Reynolds
19  Department of Emergency Medicine
20  Virginia Commonwealth University Reanimation Science Center;
21  600 East Broad St.
22  Richmond VA 23298
23  USA
24
25  PH 804 628 1942
26  FAX 804 828 4686
27  Email: psreynolds@vcu.edu






28    ***Abstract****. Transition points are common in physiological processes. However the

29    transition between normothermia and hypothermia during haemorrhagic shock has

30    rarely been systematically quantified from intensive time series data. We estimated the

31    critical transition point (CTP) and provided confidence intervals for core body

32    temperature response to acute severe haemorrhage in a conscious rat model.

33    Estimates were obtained by traditional piecewise linear regression (*broken stick* model)

34    and compared to those from the more novel *bent cable* regression. Bent cable

35    regression relaxes the assumption of an abrupt point transition, and thus allows the

36    capture of a potentially gradual transition phase; the broken stick is a special case of the

37    bent cable model. We calculated two types of confidence intervals, assuming either

38    independent or autoregressive structure for the residuals. In spite of the severity of the

39    haemorrhage, median temperature change was minor (0.8 $^{o}$C; IQR 0.57-1.31 $^{o}$C) and

40    only four of 38 rats were clinically hypothermic (core temperature < 35 $^{o}$C). However, a

41    transition could be estimated for 23 rats. Bent cable fits were superior when the

42    transition appeared to be gradual rather than abrupt. In all cases, assuming

43    independence gave incorrect uncertainty estimates of CTP. For 15 animals, neither

44    model could be fitted because of irregular temperature profiles that did not conform to

45    the assumption of a single transition. Arbitrary imposition of broken stick fits on a

46    gradual transition profile and assuming independent rather than autocorrelated error

47    may result in misleading estimates of CTP. Identification of the onset of irreversible

48    shock will require further quantification of appropriate time-dependent physiological

49    variables and their behaviour during haemorrhage.

50

51



### Introduction

Transition points are common in physiological processes. Frequently, physiological responses to specific stressors will be nonlinear, with a critical threshold occurring when the response is believed to change more or less abruptly at some level of the independent variable. This *critical transition point* (CTP) is indicative of a change in physiological state, and therefore may serve as a marker of that state change. Such changes may be relatively innocuous, such as the so-called lactate threshold (Myers & Ashley, 1997; Vachon *et al.*, 1999) or the gas exchange threshold (Kelly, 2001), which are associated with the transition from aerobic to anaerobic metabolism. In other cases, transitions indicate an undesirable pathological state change that may be impossible to reverse. One example is the transition from a state of compensated to decompensated shock, where oxygen extraction by the tissues is no longer sufficient to meet demand, resulting in ischemic metabolic insufficiency (Schumacker & Cain, 1987; Schumacker & Samsel, 1989), accumulation of oxygen debt (Dunham *et al.*, 1991; Barbee *et al.*, 2010), and increased risk of mortality. It is of great clinical importance to be able to identify early signs of shifts in regulated physiological functions that forecast the risk of abrupt pathological change. This is analogous to the problem of forecasting regime shifts in ecological systems. Unfortunately, both types of regime shift share a common problem: the only certain identifier of a regime shift threshold is to cross it (Thrush *et al.*, 2009). Nevertheless, accurate quantitative descriptions of the transition region are an essential first step to elucidating mechanisms underlying the change (Connett *et al.*, 1986).

Hypothermia is common following traumatic injury and is a significant risk factor



74  for acidosis, coagulopathy, and mortality (Peng & Bongard, 1999; Tsuei & Kearney,

75  2004; Martini, 2009). There is considerable debate as to whether hypothermia is a

76  relatively late sign in the development of shock (Peng & Bongard, 1999; Smith & Yamat,

77  2000; Beilman *et al.*, 2009), or is a complication independent of the shock process

78  (Silbergleit *et al.*, 1998). Clinical studies rarely distinguish between environmentally-

79  induced hypothermia – that is, hypothermia induced by aggravated heat loss

80  occasioned by CNS injury, intoxication, treatment interventions such as fluid

81  resuscitation, etc. (Luna *et al.*, 1987) – or hypothermia as a marker of progressive

82  metabolic dysfunction (Beilman *et al.*, 2009).  To add to the confusion, spontaneous

83  core temperature reduction occurs during haemorrhage in laboratory animal models;

84  and in this context is actually a regulated autonomic change acting to reduce metabolic

85  rate, and thus tissue oxygen demand, as oxygen delivery becomes progressively more

86  limited (Henderson *et al.*, 2000; Brown *et al.*, 2005). The extent of the temperature

87  change may also depend on the amount and rate of blood loss during haemorrhage

88  (Connett *et al.*, 1986; Wu *et al.*, 2003).  Data-driven models of body core temperature

89  over time have been used as "early warning" indicators of heat-related pathological

90  change during intense activity (Gribok *et al.*, 2006; Gribok *et al.*, 2007; Gribok *et al.*,

91  2008). However, patterns of body temperature change during haemorrhage have not

92  been examined for their utility in early prediction of the onset of decompensated shock.

93  This may be partially because core temperature reduction during haemorrhage in

94  laboratory rodents is relatively minor – less than 1$^\circ$C (Henderson *et al.*, 2000) to

95  approximately 2 $^\circ$C (Brown *et al*., 2005)  – so that detection of the transition region is

96  difficult until core temperature depression becomes pronounced.



97

98          This study examined an intensive time course of individual thermoregulatory

99   patterns during controlled severe haemorrhage in a conscious rat model. Because core

100  temperature depression was expected to be relatively subtle, the goals were first, to

101  determine if early-stage thermoregulatory transition could in fact be detected, and

102  second, to determine the most biologically meaningful and best-fitting model of that

103  transition for each individual subject. Model reliability was assessed by comparing the

104  performance of two competing piecewise regression models: the conventional *broken*

105  *stick* model (Vieth, 1989; Berman *et al.*, 1996; Toms & Lesperance, 2003) versus the

106  *bent cable* model (Toms & Lesperance, 2003; Chiu *et al.*, 2006; Chiu & Lockhart, in

107  press) as descriptors of the transition region. We also assessed the effect of correlation

108  between sequential observations on the precision of the change point estimate.

109

110  ***Methods***

111  *Ethical approval.* This study was approved in advance by the Institutional Animal Care

112  and Use Committee (IACUC) of Virginia Commonwealth University, and conforms to the

113  Public Health Service Policy on Humane Care and Use of Laboratory Animals (2002).

114

115  *Animal husbandry.* Male Long Evans rats were obtained from Harlan Laboratories

116  (Indianapolis IN) at approximately 8 weeks of age. Prior to experimentation, animals

117  were housed two to a cage and maintained at 25 (SD1) $^{o}$C and 12L:12D in the animal

118  colony. All rats had access to food (commercial rat chow) and water *ad lib*.  Rats

119  weighed an average of 314 (SD 12) g at the time of experiments.



120

121    *Core temperature and haemorrhage data.* Data reported here were collected as part of

122    a study designed to assess resuscitation strategies in a conscious rodent model that

123    best promote survival for 3 h following severe (60% total blood volume) haemorrhage

124    without conventional large-volume crystalloid support. Study guidelines were directed by

125    the Defense Advanced Research Projects Agency [DARPA] [no. BAA04-12;

126    http://www.darpa.mil/baa/baa04-12mod3.html]. Haemorrhage and resuscitation

127    procedures have been described in detail elsewhere (Reynolds *et al.*, 2007) and are

128    briefly summarized here.

129

130        Rats (n = 38) were anaesthetized with isoflurane (5% for induction, 1%–2% for

131    maintenance, balance $O_2$). During surgery core temperature was monitored with a rectal

132    probe and maintained at 36.5–37.5$^o$C with a thermostatically controlled feedback

133    heating blanket (Harvard Apparatus, Holliston, MA). Animals were catheterized and

134    implanted with an intraperitoneal temperature transponder (E-mitter; Minimitter, Bend,

135    OR) under surgically sterile conditions, then allowed to recover from anaesthesia for 30-

136    60 minutes before haemorrhage began, after they were able to both right themselves

137    and maintain a sternal position. This delay  ensured adequate time for anaesthetic gas

138    washout.  For pain control, all incisions and routing tracks were coated with topical 2%

139    lidocaine gel. An unprimed micro-osmotic pump (model 1003D; Alzet Osmotic Pumps,

140    Durect, Cupertino, CA ) filled with morphine (50 mg mLj1) was implanted

141    subcutaneously at the nape to deliver approximately 0.15 mg kg$^{-1}$ h$^{-1}$ morphine after

142    haemorrhage; animals also received a single SQ morphine injection (0.3 mg kg$^{-1}$).



143

144      Animals were haemorrhaged from the carotid catheter in three 20%-volume

145  increments at rates of 1.0, 0.5, and 0.25 mL min$^{-1}$, respectively. Blood was withdrawn by

146  a programmable syringe pump (PHD22/2000 Series, Harvard Apparatus, Holliston MA).

147  Blood draw was constrained by a minimal mean arterial pressure (MAP) threshold of 40

148  mm Hg. If MAP fell below this threshold, haemorrhage was stopped and animals were

149  then allowed to auto-resuscitate until a MAP of 40 mm Hg had been maintained for at

150  least 75 s, after which haemorrhage was resumed until target shed blood volume (SBV)

151  was achieved. Target SBV was calculated as 60% of the total blood volume (TBV),

152  which was estimated from body mass of each subject as TBV (mL) =  0.77 + 0.06

153  (mL/g) · M, where M is body mass (g) (Lee & Blaufox, 1985). Blood volumes removed

154  averaged 11.8 (SD 0.5) mL.  Animals were then resuscitated with one of four fluid

155  treatments as described previously, and monitored for 180 min or until cardiac collapse.

156  Two rats did not survive to receive the full resuscitation intervention. Animals surviving

157  to 180 min were humanely euthanized with Euthasol (pentobarbital, 0.3-0.4 mL kg$^{-1}$ or

158  100-150 mg kg$^{-1}$, IV) (Reynolds *et al.*, 2007).

159

160       Core temperature was logged by remote data collection for the duration of each

161  trial, and averaged at 15 s increments ("time steps") using an exponentially-weighted

162  moving average algorithm (VitalView® Data Acquisition System; Minimitter Inc., Bend,

163  OR). Temperature records for each animal were between 127 and 246 time steps (32 to

164  62 min) long, and for this analysis included only the haemorrhage portion of each trial.

165



166  *Models.*  Numerous quantitative or model-based approaches have been proposed for

167  identifying physiological thresholds; however the most common is to fit a piecewise

168  linear regression model (Berman *et al.*, 1996). Piecewise regression models consist of

169  two or more time trajectories of a physiological response, characterized as either lines

170  or curve segments joined at a point that typically represents the transition. The transition

171  may occur because of either a specific experimental or quasi-experimental intervention

172  (Gillings *et al.*, 1981) or a physiological state change (Schumacker & Cain, 1987; Vieth,

173  1989; Berman *et al.*, 1996; Myers & Ashley, 1997; Vachon *et al.*, 1999). Typically, the

174  process is modelled as two separate straight lines with different slopes and a common

175  intersection point at the transition point τ (Fig. 1; Vieth, 1989; Berman *et al.*, 1996); this

176  is sometimes referred to as the *broken stick* model (Toms & Lesperance, 2003).

177

178  The broken stick model describes a single time course of core temperature data

179  $y_t$ as:

180
$$y_t = \begin{cases} \beta_0 + \beta_1 t + \varepsilon_t & \text{for } t < \tau \\ \beta_0 + \beta_1 t + \beta_2(t - \tau) + \varepsilon_t & \text{for } t \geq \tau \end{cases} \qquad (1)$$

181  where $\beta_0$ is the baseline core temperature, $\beta_1$ is the rate of temperature change before

182  the transition point τ, $\beta_1 + \beta_2$ is the rate of temperature change after the transition point

183  τ, and $\varepsilon_t$ is the regression error, or residual, at time t. The model is constrained by the

184  intersection of lines at τ; this point is unknown and must be estimated from data. Here,

185  the critical transition point CTP is equal to τ.

186

187  Although convenient and easy to implement, estimates derived from the broken



188    stick model may be compromised by the scientifically uncritical assumption of an abrupt

189    transition point. Not only may this model lack theoretical justification, in many cases it

190    may not even resemble the actual data (Jones & Handcock, 1991; Routledge, 1991;

191    Myers & Ashley, 1997).  For many transition series, a more realistic model may be the

192    *bent cable* regression model (Chiu *et al.*, 2006; Chiu & Lockhart, in press). Here, the

193    data are described in terms of two linear segments as before, but the transition between

194    the two segments is modelled as a quadratic bend (Tishler & Zhang, 1981), such that

195

$$y_t = \begin{cases} \beta_0 + \beta_1 t + \varepsilon_t & \text{for } t < \tau - \gamma \\ \beta_0 + \beta_1 t + \beta_2 (t - \tau + \gamma)^2 / 4\gamma + \varepsilon_t & \text{for } \tau - \gamma \le t \le \tau + \gamma \\ \beta_0 + \beta_1 t + \beta_2 (t - \tau) + \varepsilon_t & \text{for } t > \tau + \gamma \end{cases} \quad (2)$$

196    (Fig. 1). Here $\gamma$ is a non-negative parameter that determines the width of the quadratic

197    transition zone that is centred at $\tau$ and ranges from the lower bound at $(\tau - \gamma)$ to the

198    upper bound at $(\tau + \gamma)$. Unlike the broken stick model, the bent cable model is smooth

199    with no obvious change point. Instead, the transition is more accurately characterized

200    as the range of time over which the trajectory has a positive slope, followed by the time

201    range where core temperature declines and the trajectory has a negative slope.

202    Therefore CTP occurs where the slope of the trajectory inside the quadratic bend is

203    equal to 0, so that the CTP is $[\tau - \gamma - (2\beta_1\gamma/\beta_2)]$. However if $\gamma = 0$, the model reduces to

204    Eqn. 1 and CTP equals $\tau$; therefore, the broken stick model is a special case of the bent

205    cable model (Chiu *et al.*, 2006).

206

207    *Autocorrelation*. One key assumption of common regression-type models is that the

208    residuals $\varepsilon_t$s are independent of each other. However, with time series data, these

209    residuals are usually correlated over time; the correlation between sequential



210    observations is $\rho$. If this autocorrelation is ignored (that is, $\rho$ is naively assumed to be

211    equal to zero), width of the confidence intervals will be misleadingly narrow. In practical

212    terms, when the value of one measurement influences the determination of another,

213    then the information available from these two measurements combined is less than that

214    obtained from two independently-observed measurements. Therefore, this

215    autocorrelation must be accounted for in the model.

216

217        When time intervals are equally spaced, an autoregressive (AR) structure for the

218    residuals is often adequate; AR models are appropriate when it is reasonable to

219    assume that any currently-observed value of the series depends on immediate past

220    values, plus a random error component. Thus, the autoregressive component is a linear

221    regression of the current value of the residual at time t ($\varepsilon_t$), where the independent

222    variables are the $p$ previous values of the time series $\varepsilon_{t-1}$, $\varepsilon_{t-2}$, ... , $\varepsilon_{t-p}$. The quantity $p$ is

223    the "order" of the autoregression, and is a measure of the amount of memory in the

224    system. For example, an autoregressive series with $p = 3$ means that the current value

225    of the series is dependent on its previous $p = 3$ lagged values, and is said to have an

226    AR(3) structure. If there is no autocorrelation between observations (i.e. the series is

227    white noise) then $p = 0$.  In general, the autocorrelation between sequential

228    observations $y_t$ is expressed in terms of autocorrelated residuals as

229            $$\varepsilon_t = \varphi_1 \cdot \varepsilon_{t-1} + \varphi_2 \cdot \varepsilon_{t-2} + ... + \varphi_p \cdot \varepsilon_{t-p} + \delta_t \qquad\qquad (3)$$

230    where $\varphi_1,...,\varphi_p$ are the AR coefficients to be estimated, and $\delta_t$s are the uncorrelated

231    autoregression errors (or white noise) with variance $\sigma^2$. If $p = 0$, then $\varepsilon_t$ reduces to white

232    noise $\delta_t$ (Box *et al.*, 1994). Based on asymptotic statistical properties of the conditional



233  likelihood function (Chiu & Lockhart, in press), the 95% confidence interval for CTP can

234  be approximated as the estimated value of CTP $\pm$ [1.96 $\cdot(\xi'\, \boldsymbol{V}\, \xi)^{1/2}$], where $\xi'$ is the row

235  vector [ 0, -2 $\cdot$ $\gamma/\beta_2$, 2$\cdot\beta_1 \cdot \gamma/\beta_2^2$, 1, -(2$\cdot\beta_1 + \beta_2)/\beta_2$ ], and $\boldsymbol{V}$ is the 5 x 5 asymptotic

236  covariance matrix for the vector of estimated bent cable regression coefficients,

237  obtained from Eqn. 2.

238

239  *Model fitting.* The model fitting process consisted of three steps. First, preliminary

240  structural model fits were obtained separately for each subject. Because the broken

241  stick model is a special case of the bent cable model, the broken stick fit was obtained

242  by constraining $\gamma$ = 0; to obtain the bent cable fit, this constraint was removed. In this

243  preliminary model a "naïve" fit was obtained by assuming independent errors (i.e. $p$ = 0).

244  Second, we obtained the residuals from this preliminary model and applied standard

245  autocorrelation diagnostics (ACF, PACF) to determine the order $p$ of the autoregression

246  (Box *et al.*, 1994); this value for $p$ was substituted in Eqn. 3. Finally parameters for the

247  full model (i.e. $\sigma^2$ and all unknown coefficients from Eqn. 2 and 3) were simultaneously

248  estimated by conditional maximum likelihood (Chiu & Lockhart, in press). Because there

249  is no analytical solution, goodness of fit was assessed by visual inspection of the fitted

250  curve and residual diagnostics (Fig. 2). If the tentatively-entertained model is correct,

251  then plots of $\varepsilon_t$ over time should exhibit mean 0 and constant variance with no

252  heteroscedasticity and no departure from randomness (Box *et al.*, 1994). Six data series

253  required minor trimming of data ($\leq$ 10 data points) on either end of the sample period to

254  allow model convergence and sensible estimation of model parameters.  Model fitting

255  was performed using the statistical computing software R with the *bentcableAR*



256    package; code is available online at *http://www.r-project.org.*

257

### *Results*

259          In spite of the severity of the haemorrhage protocol, only four of 38 rats in this

260    study showed signs of clinically mild hypothermia [core temperature < 35 $^{o}$C; (Beilman

261    *et al.*, 2009)] at the end of haemorrhage; four were borderline (core temperature

262    between 35-36 $^{o}$C). Temperature of the remaining animals averaged 37.13 (SD 0.55) $^{o}$C

263    at the end of haemorrhage. The median difference between peak temperature and that

264    recorded at the end of haemorrhage before fluid resuscitation ($\Delta$T) was 0.8 $^{o}$C (IQR

265    0.57-1.31 $^{o}$C). However, although most animals were technically normothermic following

266    haemorrhage, data for 23 animals showed a detectable transition in core temperature

267    trajectory during haemorrhage (Table 1).  For these subjects, there was reasonable

268    within-subject agreement in estimates of the transition point. Estimated CTP averaged

269    17.5 and 16.1 min respectively for the broken stick and bent cable models with

270    independent errors, and 17.7 and 15.5 min with autoregressive errors. For 9 subjects,

271    the bent cable-autoregressive model was a better fit than the broken stick–

272    autoregressive model (Fig 2.A); discrepancies between models in estimates of CTP

273    were as large as 4 min in some cases.

274

275          When the estimate of $\gamma$ was exactly or approximately zero, two technical

276    complications resulted. First, estimates of CTP from the bent cable model (Eqn. 2) were

277    often (although not always) virtually identical to the broken stick model (Eqn. 1; Fig.

278    2.B); however because $\gamma$ = 0, the resulting confidence intervals for the estimate were



degenerate. A different type of degeneracy resulted from bent cable fits where the slope

never changed sign, so that CTP is undefined and the corresponding confidence

intervals could not be computed. For subjects where CTP estimates were virtually

identical between methods, visual inspection of graphics and residuals (Fig.3)

suggested the most appropriate model fit.

For all 23 series, the data showed statistically significant correlation. The

autoregressive order $p$ was between 2 and 4 for 19 cases, and between 5 and 7 for four

cases (Table 1), indicating a time-dependent "persistence" (Santer *et al.*, 2008), or lag,

of between 30 sec to almost 2 min. When autoregressive error was incorporated into

the full model, confidence intervals from broken stick fits were substantially narrower

than those from bent cable fits (Table 1).  Estimates of CTP obtained from broken stick

models appeared to be sensitive to the accounting of autocorrelation in the error

structure. This was suggested by the observation that in many cases CTP and

corresponding confidence intervals differed considerably between models incorporating

autoregressive errors (that is, $p > 0$) compared to those assuming independence ($p = 0$;

Table 1).

For 15 cases, no single CTP could be clearly identified because temperature

profiles were too irregular or oscillatory to allow a reasonable fit by either model. Two

examples are shown in Fig. 4.  In the first case, relatively large-scale amplitude

oscillations (0.5 $^{o}$C) in core temperature began with the onset of haemorrhage and

continued to haemorrhage cessation. In the second case, an initial 1$^{o}$C drop in

temperature beginning approximately 6 minutes into the haemorrhage was followed by



302    damped oscillations around what may be a new reduced set-point (Henderson *et al.*,

303    2000; Brown *et al.*, 2005).

304

305    **DISCUSSION**

306         In this study we evaluated time-dependent patterns of body temperature

307    regulation during haemorrhage in rats, with models assuming the existence of a defined

308    transition between relative normothermia and the onset of core temperature depression.

309    Methods were based on a regression model incorporating the novel use of bent cable

310    regression (with its relaxed assumptions as to the smoothness of the transition region),

311    together with parametric time-series models (to correct estimation problems associated

312    with residual autocorrelation). Although only four animals could be considered

313    technically hypothermic at the end of haemorrhage, threshold models could be fitted to

314    data for approximately two-thirds of the subjects in this study. These data showed that

315    even subtle changes in core temperature can be detected during intensive monitoring.

316

317         Shifts in physiological regulatory steady state as a result of some stressor are

318    often described by thresholds (Vieth, 1989; Berman *et al.*, 1996), critical points

319    (Schumacker & Cain, 1987), deflection points (Vachon *et al.*, 1999), or change points

320    (Jones & Handcock, 1991; Berman *et al.*, 1996; Kelly, 2001). Numerous studies have

321    used piecewise regression models to estimate these transitions, e.g. (Schumacker &

322    Cain, 1987; Vieth, 1989; Jones & Handcock, 1991; Samsel *et al.*, 1991; Myers &

323    Ashley, 1997; Vachon *et al.*, 1999; Kelly, 2001). Many previous approaches have

324    assumed that the transition is abrupt and clearly demarcated, and tacitly assumed that



325   the data on each subject consisted of independent observations (because potential

326   autocorrelation was not explicitly accounted for in the models). In this study, we

327   demonstrated that ignoring the shape of the transition region could affect estimates of

328   the transition time. Uncritical use of $\tau$ estimated from the broken stick model could

329   impede timely recognition of the transition recognition if an abrupt transition is assumed,

330   but the transition is actually gradual. Second, we showed that, if the correlation between

331   sequential observations is ignored, calculated confidence intervals are often

332   misleadingly narrow, giving a false impression of the precision of the estimates. In this

333   study, the assumption of independent and identically-distributed errors was not valid for

334   any subject. Although theoretical considerations suggest (and our data show) that

335   ignoring autocorrelation will in many cases yield estimates of $\tau$ and CTP consistent with

336   those obtained by incorporating autocorrelation in the final model, precision is greatly

337   affected by failure to incorporate appropriate error structure. Consequently, we consider

338   that autocorrelation between observations should be incorporated as part of the model,

339   and conditional maximum likelihood estimation applied for parameter estimation. If

340   within-subject autocorrelation is negligible or absent, the parameters of the broken stick

341   model can be estimated for each subject by using any standard nonlinear estimation

342   procedure, and confidence intervals for $\tau$ obtained using asymptotic standard errors.

343   However, for long data series sampled at frequent short time intervals, autocorrelation is

344   unlikely to be trivial, and these standard errors will not be reliable.

345

346         When autoregressive errors were incorporated into the models, confidence

347   intervals for CTP derived from broken stick fits were narrower than those from bent



348    cable fits. This result does not suggest that estimates derived from broken stick models

349    are inherently less uncertain than the bent cable fits; rather this apparent "precision" is

350    most likely an artefact of the more restrictive assumptions associated with the broken

351    stick model. Specifically, when an abrupt breakpoint estimate is forced on a relatively

352    smooth or gradual profile, the range of time steps required to bracket that estimate will

353    be necessarily much smaller than those required for a smooth curve. In these cases,

354    the broken stick fit could result in estimates of CTP and confidence intervals that poorly

355    represent the observed profile.

356

357        Typically, effective fitting of autoregressive models requires at least a moderately

358    long series; in general at least 50 observations are recommended (Box *et al.*, 1994;

359    Chatfield, 1996).  In practice, this should be feasible if subjects are appropriately

360    instrumented, and physiological data can be obtained as a single long intensive time

361    series. In this study, data were sampled at extremely short (15 s) time intervals so that it

362    was possible to capture the transition region and visually assess the adequacy of each

363    model fit to the transition point.

364

365        To arrive at the actual "best" solution for a transition point requires a great deal of

366    care, essentially consisting of an exhaustive search among the numerous "near-best"

367    solutions for identifying the actual "best" solution. In a few cases, truncation or trimming

368    a small number of data points at the end of the core temperature profile will allow

369    sensible model fits. In this study, trimming was employed for six subject profiles (Table

370    1). However, model estimates may not be possible if the physiological profile for a given



371    subject does not exhibit a structure that resembles either biphasic model described

372    here. For 15 cases examined in this study, no thermoregulatory transition could be

373    clearly distinguished, and neither model provided a sensible fit. In these cases,

374    mathematical irregularities of the regression arising from low-order differentiability will

375    prevent estimation of model parameters (Chiu *et al.*, 2002, 2005, 2006). These may

376    occur because of excessive oscillations in the data series (Fig. 4). In these cases, even

377    a statistical "best" fit may be nonsensical in practice because the postulated model does

378    not provide an adequate description of the series behaviour. Oscillations in body

379    temperature occurred either almost immediately after haemorrhage onset (< 20% blood

380    volume removed) or appeared to be sustained around what may have been a new set-

381    point (Brown *et al.*, 2005) following more extensive blood loss (30-40% total blood

382    volume; Fig. 4). Oscillatory instabilities in otherwise periodic (Mackey & Glass, 1977) or

383    stable (Stark & Baker, 1959) physiological systems are common in many disease

384    conditions (Mackey & Glass, 1977). Unstable oscillations in body temperature with

385    haemorrhage may result from enhanced activity of, and feedback from, the sympathetic

386    system, as is the case for arterial pressure oscillations during haemorrhage (Hosomi,

387    1978). Thus, uncritical use of any biphasic regression model is not recommended

388    without visual inspection of the data beforehand.

389

390        If a transition region can be determined to exist based on external criteria, some

391    otherwise intractable problems of individual model fits might be overcome by combining

392    data for all individuals using a single longitudinal mixed-effects model. This will be

393    appropriate if estimates of population characteristics, rather than subject-specific



394    quantities, are desired. For example, to compute the subject-specific transition point for

395    each individual profile in this study there were five regression parameters ($\beta_0$, $\beta_1$, $\beta_2$, $\gamma$,

396    $\tau$), plus $p$ autoregressive parameters and an error variance to be estimated. However, if

397    all $n$ individual profiles can be considered collectively as a sample from a population, a

398    single population cable profile with $(5 + p)$ unknown parameters, plus a handful of

399    covariance parameters, could be estimated from the pooled data for all $n$ subjects; this

400    leads to a single estimate of the population transition point. The two approaches differ

401    because random between-subject variation is incorporated in the mixed-effects model

402    (Khan *et al.*, 2009; Khan *et al.*, submitted), and could be useful for other physiological

403    models.

404

405    Our method of estimating critical transition regions for individual subjects will be

406    relevant to the study of virtually any dynamic physiological process characterized by a

407    rate change in physiological state. In general, bent cable regression is a viable and

408    preferred generalization of classic piecewise linear regression that allows for shape

409    variability in the transition between phases. We recommend its use for situations where

410    flexibility is desired, either because the underlying physiological mechanisms are not

411    well understood, or where there is no strong supporting theory or evidence for a model

412    of abrupt change in the transition between physiological phases. Because arbitrary

413    imposition of broken stick fits on a gradual transition profile may result in very

414    misleading estimates, broken stick fits should be avoided unless preliminary model

415    estimation results in values of $\gamma$ that are near zero, indicating that a broken stick model

416    is appropriate. Second, dependence between sequential observations cannot be



417  ignored. This is especially apparent in cases where a broken stick model is appropriate,

418  as this model appears very sensitive to autocorrelation.

419

420       Early warning of a transition to physiological instability is highly desirable so that

421  appropriate interventions can be performed before the subject is beyond help.

422  Physiological monitoring devices have greatly improved in recent years so that

423  extremely high-frequency sampling of variety of physiological variables is possible,

424  enabling the generation of highly concentrated time series data over very short periods

425  of time. This will allow the exploration of the behaviour of a wide variety of potential

426  indicator variables. However identifying which of these variables can actually indicate

427  incipient transition before the transition actually occurs is still difficult. Determination of

428  such indicators will require both robust mathematical modelling and, more important,

429  collection of long-term data series under carefully controlled conditions. Accurate

430  estimates of time to this transition may serve ultimately to improve trauma management

431  through better and earlier detection of shock and optimization of the timing and extent of

432  therapeutic interventions.

433

434  ***Acknowledgements***


435  We thank M. Skaflen for superb technical assistance, and N. J. White and R. W. Barbee

436  for useful comments on the manuscript. The statistical work of this article was funded by

437  Discovery Grant #RGPIN261806-05 (to GSC) from the Natural Sciences and

438  Engineering Research Council (NSERC) of Canada. The haemorrhage-resuscitation

439  studies were funded by the Department of Defense DARPA Surviving Blood Loss




440    Program, contract # N66001-02-C-8052 (to R. W. Barbee) and the Department of

441    Emergency Medicine.

442

### References

444

Table 1. Estimates of the critical transition point CTP of body temperature for 23 rats during severe controlled haemorrhage.

| Subject | Independent errors ρ = 0 | | | | p | Autoregressive error | | | | γ |
| | Broken stick | CI | Bent cable | CI | | Broken stick | CI | Bent cable | CI | |
| --- | --- | --- | --- | --- | --- | --- | --- | --- | --- | --- |
| 1 | 24.71 | 1.65 | 24.30 | 2.29 | 2 | 23.63 | 0.87 | 23.63 | . | 0.00 |
| 2 | 15.87 | 1.23 | 14.80 | 0.69 | 3 | 24.25 | 0.53 | 14.80 | 2.57 | 10.31 |
| 3 | 19.10 | 1.38 | 19.02 | 1.49 | 7 | 17.86 | 0.50 | 17.86 | . | 0.00 |
| 4 | 19.12 | 0.90 | 15.92 | 0.78 | 5 | 16.38 | 0.44 | 15.82 | 1.24 | 9.69 |
| 5* | 20.10 | 1.01 | 18.55 | 1.06 | 2 | 21.41 | 0.24 | 18.09 | 2.06 | 6.12 |
| 6 | 15.35 | 1.66 | 17.42 | 0.73 | 4 | 15.00 | 0.79 | 17.60 | 3.07 | 10.31 |
| 7* | 15.43 | 0.36 | 13.69 | 0.58 | 6 | 16.92 | 0.10 | 14.84 | 1.22 | 1.28 |
| 8* | 6.19 | 1.04 | 6.25 | 1.50 | 2 | 5.55 | 0.47 | 5.57 | 0.82 | 0.06 |
| 9* | 9.70 | 0.80 | 9.81 | 0.58 | 3 | 10.24 | 0.42 | 9.92 | 0.94 | 3.91 |
| 10 | 12.18 | 0.68 | 13.11 | 1.77 | 4 | 11.55 | 0.55 | 13.51 | 3.86 | 3.04 |
| 11 | 15.63 | 0.83 | 12.41 | 0.45 | 3 | 23.28 | 0.39 | 12.43 | 1.41 | 9.69 |
| 12 | 4.91 | 0.87 | 5.89 | 1.65 | 4 | 6.63 | 0.43 | 6.64 | . | 0.00 |
| 13* | 15.06 | 0.44 | 14.20 | 0.80 | 3 | 13.41 | 0.24 | 13.94 | 1.54 | 1.95 |
| 14 | 32.43 | 2.06 | . | . | 2 | 38.81 | 0.46 | . | . | 0.09 |
| 15 | 17.00 | 6.50 | 16.96 | . | 3 | 17.75 | 1.44 | 17.75 | . | 0.00 |
| 16 | 16.68 | 4.65 | 16.64 | . | 3 | 22.50 | 1.54 | 18.49 | . | 0.00 |
| 17 | 34.37 | 1.82 | . | . | 3 | 30.75 | 0.65 | . | . | 0.02 |
| 18 | 20.87 | 0.99 | . | . | 4 | 23.27 | 0.88 | 23.27 | . | 0.00 |
| 19 | 12.52 | 1.05 | 12.50 | . | 3 | 12.69 | 0.34 | 12.69 | . | 0.00 |
| 20 | 36.92 | 1.16 | . | . | 5 | 38.90 | 0.14 | . | . | 0.24 |
| 21 | 20.15 | 0.77 | 20.15 | . | 2 | 15.50 | 0.33 | 15.50 | . | 0.00 |
| 22* | 11.60 | 2.05 | 11.00 | 6.12 | 3 | 19.75 | 0.37 | 10.78 | 2.61 | 0.59 |
| 23 | 24.00 | 2.64 | . | . | 3 | 24.06 | 1.12 | 24.03 | . | 0.00 |



Values are in min. For the broken stick model, CTP = $\tau$ (Eqn. 1), and CTP = $[\tau - \gamma - (2\beta_1\gamma/\beta_2)]$ for the bent cable model

(Eqn. 2). CI is the width (min) of the 95% confidence interval for the CTP, assuming either independent (p = 0) or

autoregressive (p > 0) errors; order of the autoregressive function is given by p. The curvature of the temperature profile

during transition is represented by $\gamma$. When the estimate for $\gamma$ is very close to 0, the CI for CTP does not exist, and is

indicated as (.). Asterisks (*) indicated series that had minor trimming of the terminal portion of the series to allow model

convergence and better fit.

**FIGURE LEGENDS**

Fig 1. Schematic plot of bent cable (solid line) and broken stick (dashed line) regressions with identical values of $\beta_0$, $\beta_1$, $\beta_2$ and $\tau$ (see text for details). The transition point $\tau$ for the broken stick regression is the intersection of two straight lines, and is identical to the critical transition point CTP. The CTP for the bent cable regression is given by $\tau - \gamma - (2 \cdot \beta_1 \cdot \gamma / \beta_2)$, but converges to $\tau$ as $\gamma$ approaches 0.

Fig. 2. Representative plots of core temperature profiles for individual rats showing the fitted bent cable (dashed black line) and broken stick (solid gray line) models overlaid on the observed data. Solid bars indicate duration of each haemorrhage increment.

A. Plot of core temperature profile for rat 11. The bent cable fit is clearly a better representation of the transition than the broken stick model; estimated $\gamma$ is non-zero.

B. Plot of core temperature profile for rat 12. Here both the bent cable and the broken stick models give identical fits; estimated $\gamma$ for the bent cable model is zero.

Fig. 3. Plot of core temperature profile, model fits, and diagnostics for rat 9.

A. Plot of core temperature profile showing the fitted bent cable (dashed black line) and broken stick (solid grey line) models overlaid on the observed data. Solid bars indicate duration of each haemorrhage increment. Although estimates of the time to transition are similar, the transition region of the bent cable fit approximates the observed data more closely than the abrupt corner of the



broken stick fit; estimated $\gamma$ is non-zero.

B. Plot of estimated residuals $\delta_i$s  with time obtained from the bent cable fit. The scatterplot shows a random pattern, indicating an adequate fit.

Fig. 4. Core temperature profiles for two representative subjects for which neither model could be fitted. Solid bars indicate duration of each haemorrhage increment.

A. Note relatively large amplitude (0.5 $^o$C) oscillations in core temperature beginning with haemorrhage onset

B. Damped oscillation following a 1$^o$C reduction to new temperature set-point.

Fig. 1

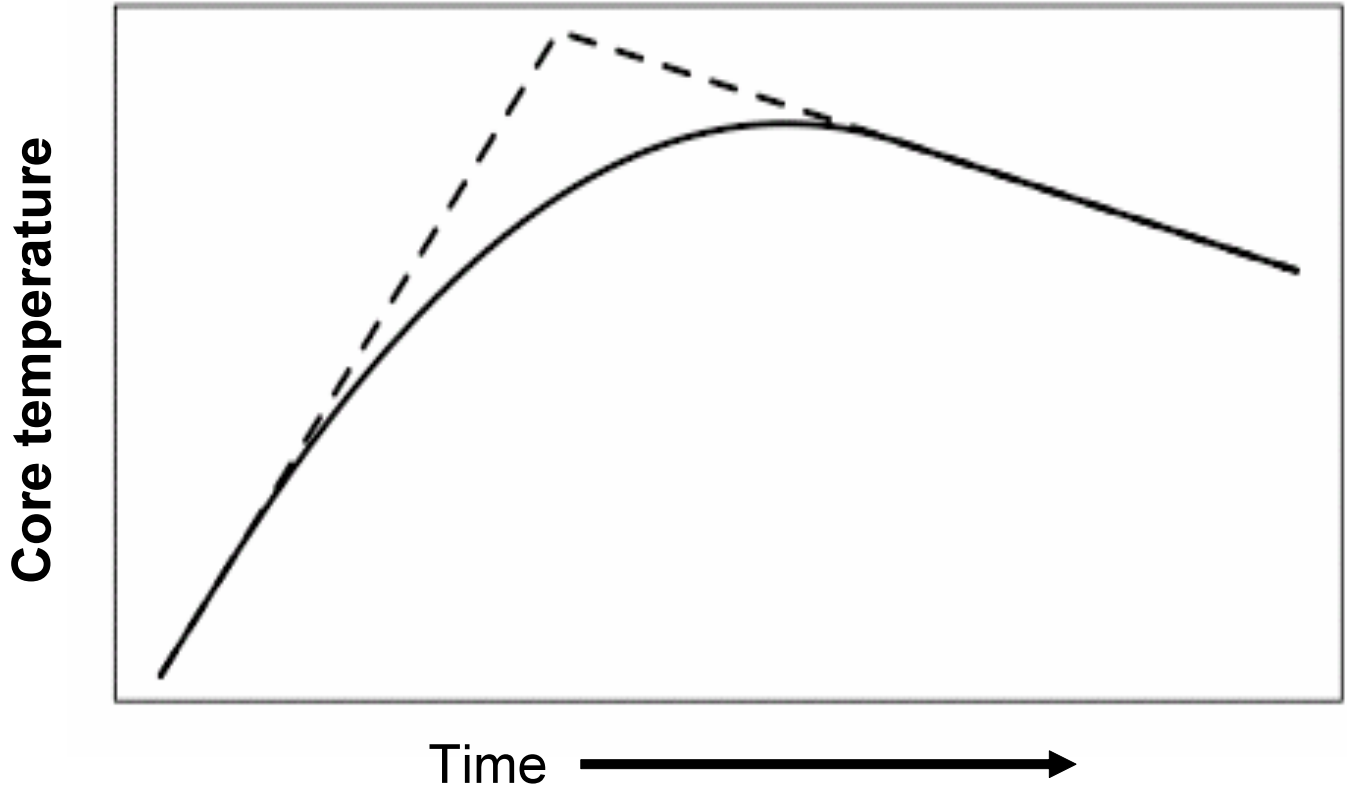

Fig. 2

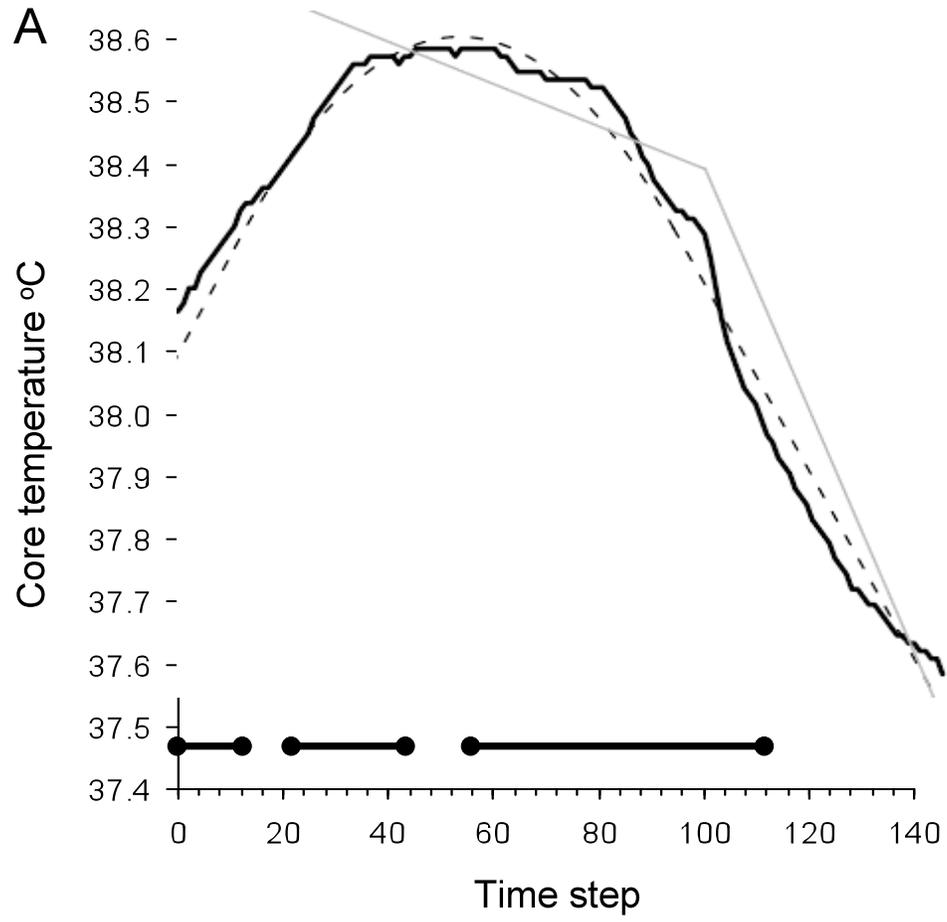

A

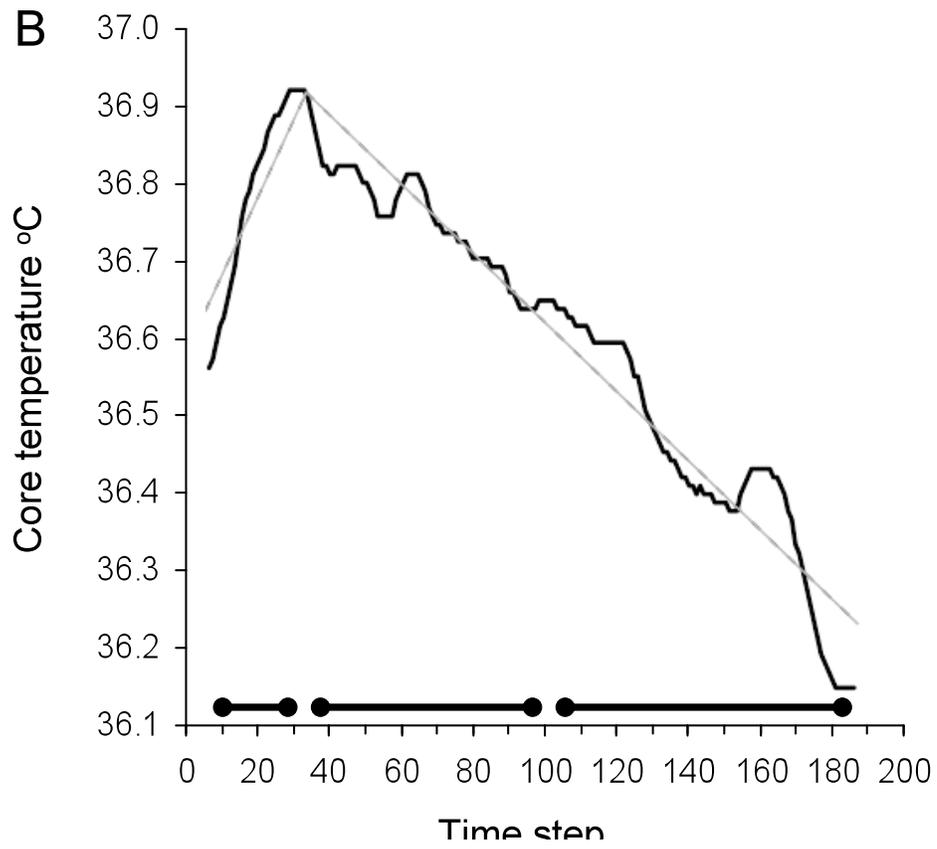

B



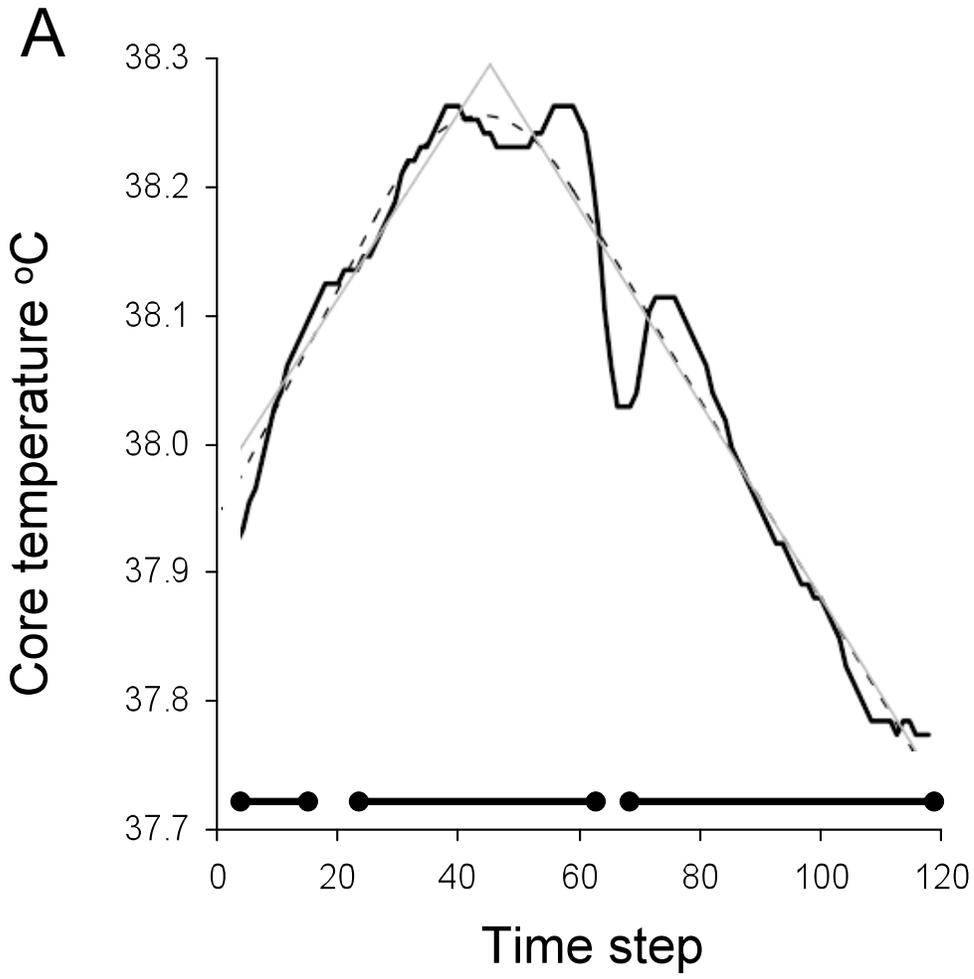

A

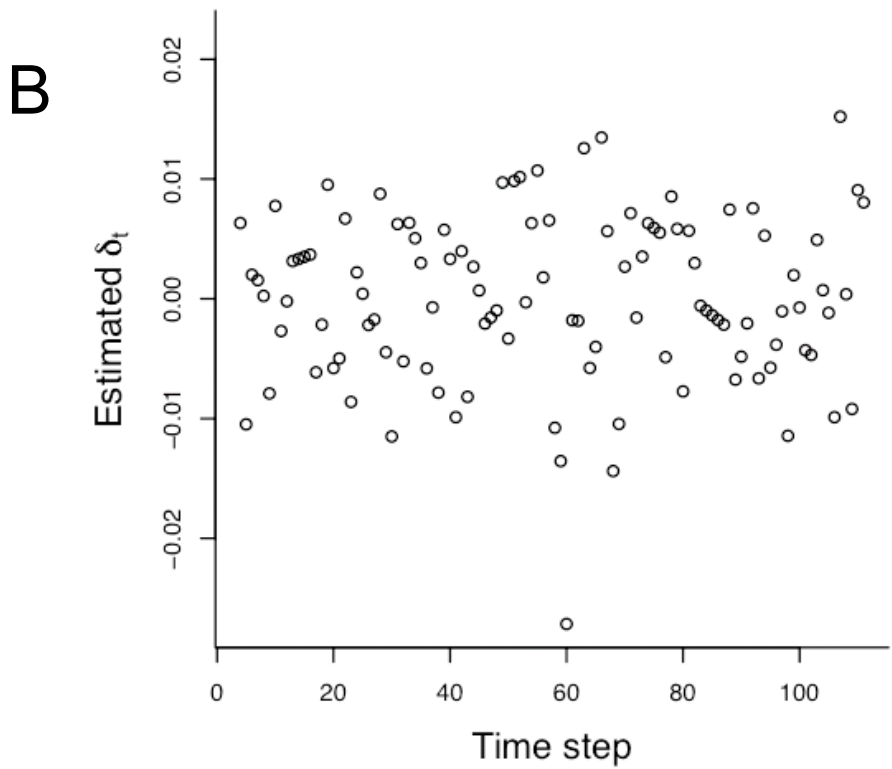

B

Fig. 4

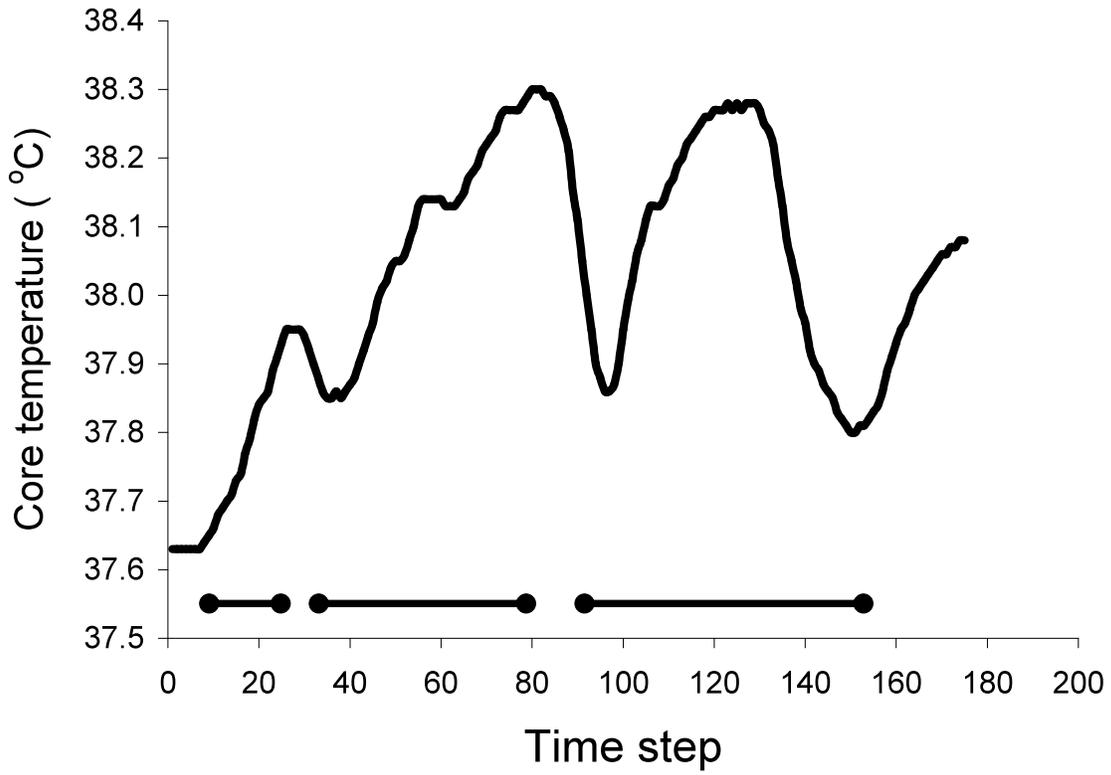

A

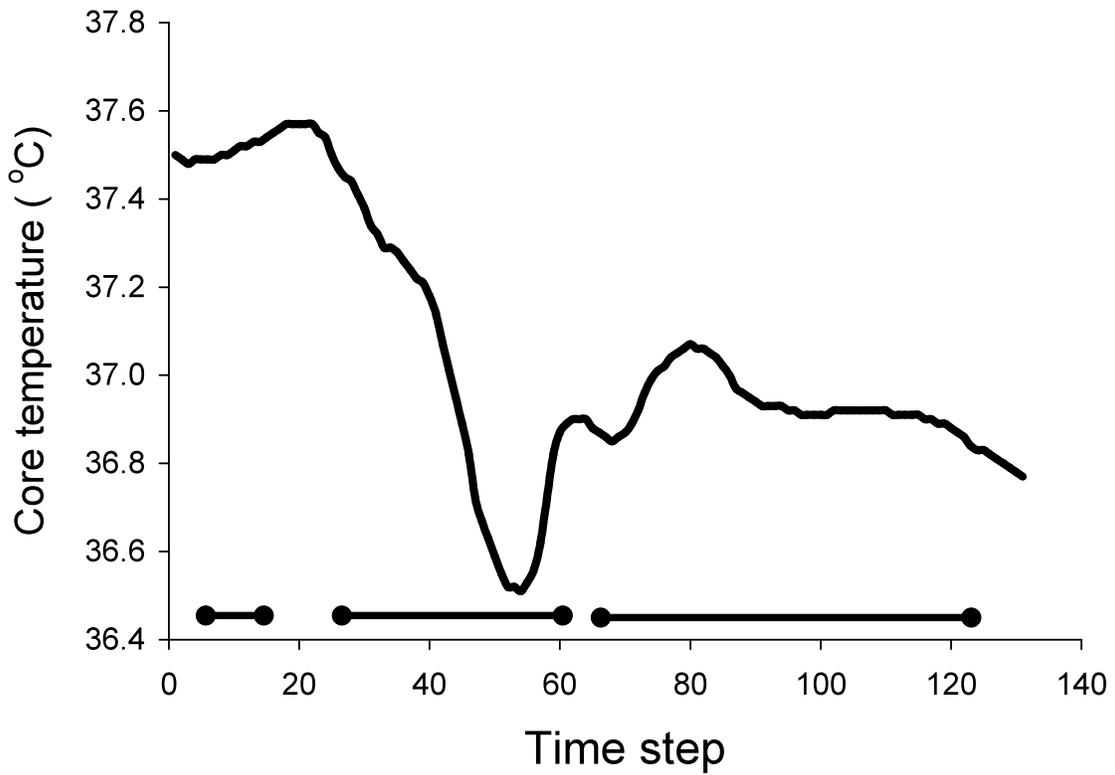

B